\documentclass[a4paper,fleqn]{cas-sc}

\usepackage[square,sort,comma,numbers, longnamesfirst]{natbib}
\usepackage[anythingbreaks]{breakurl}
\usepackage{amsthm}
\usepackage{amsmath}
\usepackage{amsfonts}
\usepackage{amssymb}
\usepackage{amsfonts}
\usepackage{epsfig}
\usepackage{bbm}
\usepackage{todonotes}
\usepackage{soul}

\usepackage{nicefrac}
\usepackage{booktabs}
\usepackage{siunitx}
\usepackage{subcaption}

\usepackage{placeins}
\usepackage{flafter}
\usepackage{caption}
\usepackage{float} 
\usepackage{rotating}
\usepackage{array}

\newcommand{\tens}[1]{\boldsymbol{\mathrm{#1}}}

\newcommand{\ICdev}{\mathrm{I}_{\tens{\bar C}}}
\newcommand{\IICdev}{\mathrm{II}_{\tens{\bar C}}}
\newcommand{\IIICdev}{\mathrm{III}_{\tens{\bar C}}}

\newcommand{\IC}{\mathrm{I}_{\tens{C}}}
\newcommand{\IIC}{\mathrm{II}_{\tens{C}}}
\newcommand{\IIIC}{\mathrm{III}_{\tens{C}}}

\DeclareMathOperator{\erf}{erf}

\graphicspath{{./images/}}

\def\tsc#1{\csdef{#1}{\textsc{\lowercase{#1}}\xspace}}
\tsc{WGM}
\tsc{QE}

\begin{document}
\let\WriteBookmarks\relax
\def\floatpagepagefraction{1}
\def\textpagefraction{.001}

\shorttitle{Rediscovering the Mullins Effect With DSR}    

\shortauthors{R. Abdusalamov, J. Weise, M. Itskov}  

\title [mode = title]{Rediscovering the Mullins Effect With Deep Symbolic Regression}

%

\author[1]{Rasul Abdusalamov}[]

\cormark[1]


\ead{abdusalamov@km.rwth-aachen.de}

\ead[url]{https://www.km.rwth-aachen.de}


\affiliation[1]{organization={Department of Continuum Mechanics, RWTH Aachen University},
	addressline={Eilfschornsteinstr. 18}, 
	city={Aachen},
	postcode={52062}, 
	state={NRW},
	country={Germany}}

\author[1]{Jendrik Weise}[]


\ead{weise@km.rwth-aachen.de}

\ead[url]{https://www.km.rwth-aachen.de}



\author[1]{Mikhail Itskov}[]

\ead{itskov@km.rwth-aachen.de}

\ead[url]{https://www.km.rwth-aachen.de}


\cortext[1]{Corresponding author}

\fntext[1]{}


\begin{abstract}
The Mullins effect represents a softening phenomenon observed in rubber-like materials and soft biological tissues.
It is usually accompanied by many other inelastic effects like for example residual strain and induced anisotropy.
In spite of the long term research and many material models proposed in literature, accurate modeling and prediction of this complex phenomenon still remain a challenging task.  \\
In this work, we present a novel approach using deep symbolic regression (DSR) to generate material models describing the Mullins effect in the context of nearly incompressible hyperelastic materials. The two step framework first identifies a strain energy function describing the primary loading. Subsequently, a damage function  characterizing the softening behavior under  cyclic loading is identified. The efficiency of the proposed approach is demonstrated through benchmark tests using the generalized the Mooney-Rivlin and the Ogden-Roxburgh model. The generalizability and robustness of the presented framework are thoroughly studied. In addition, the proposed methodology is extensively validated on a temperature-dependent data set, which  demonstrates its versatile and reliable performance.
\end{abstract}


\begin{highlights}
\item Deep symbolic regression is applied to automatically generate accurate analytical models capturing the Mullins effect in elastomers
\item Highly specific damage models accurately representing complex characteristics of the Mullins effect, including temperature-dependent effects are generated
\item Validation of the framework with multiple data sets, including temperature-dependent experimental results
\item Robustness and generalizability of the proposed framework under sparse data conditions are demonstrated
\end{highlights}

\maketitle
\section{Introduction}
\label{sec:Introduction}

Elastomers and in particular reinforced rubbers usually exhibit softening behavior appearing after very first loading. This effect was first documented in \cite{Mullins1969} and is known as the Mullins effect. It becomes
more pronounced with the increasing amount of filler particles as for example carbon black or silica and is usually accompanied by many other inelastic effects such as residual strain and induced anisotropy (see e.g. \cite{itskov2006experimental}). According to many experimental studies, all such effects depend on the actual strain level, strain rate and loading history. Moreover, the Mullins effect is very persistent and does not disappear after a long relaxation time. However, it can be reversed by exposure to high temperatures in vacuum \cite{LarabaAbbes2003}.

Despite its widespread appearance and importance, the physical source of the Mullins effect is still controversial, see e.g. \cite{Plagge2019} and references therein. There are several physical interpretations of the Mullins effect such as  detachment of polymer molecules from the filler interface, molecular slippage, rupture of filler clusters, just to mention a few. Some of these interpretations serve as a basis for micro-mechanically motivated material models of the Mullins effect.

One of the first phenomenological models of the idealized isotropic Mullins effect was proposed by Ogden and Roxburgh \cite{OgdenRoxburgh1999}. This pseudo-elastic model includes a damage parameter represented as function of the maximal value of the strain energy density previously reached under tensile loading. In \cite{Qi2004} two phases of a soft domain and a hard domain were considered. The transition from the hard to soft domain is controlled by a state variable. An updated version of the Ogden-Roxburgh model \cite{Dorfmann2004} used an alternative dissipation function to represent stress softening effects and permanent set. Another phenomenological model of the Mullins effect including permanent set and induced anisotropy was proposed in \cite{Diani2006}.
A thermodynamically consistent phenomenological model of the anisotropic Mullins effect including permanent set was further formulated in terms of principal stretches \cite{itskov2010}. In order to describe the anisotropic Mullins effect in carbon black filled rubbers a micro-mechanically motivated approach was proposed in \cite{Dargazany2009}. Khi\^{e}m and Itskov have presented an averaging based tube model for unfilled and filled elastomers \cite{Khiem2017}.
 
Many of the proposed constitutive models are biased often focusing only on some specific material effects and disregarding further significant ones. Their formulation requires expert knowledge and is very time consuming.
Besides, the parameter identification and fitting to experimental data are often expensive from the computational
point of view. Such drawbacks can however be avoided by applying modern methods of machine learning to the constitutive modeling and especially to the complex inelastic behavior of rubber-like materials. One of the first applications was reported by Shen et al.\cite{Shen2004} who created a neural network based constitutive model for rubber. A brilliant idea was to integrate continuum mechanics into the design of the neural network to provide a better understanding of the underlying physical phenomena. This was done by designing the input space of the feedforward neural network as an invariant space \cite{Shen2004}. Linka et al. extended this approach and created invariant-based constitutive artificial neural networks (CANNs) \cite{Linka2021}. Klein et al. introduced ANNs incorporating the polyconvexity condition \cite{Klein2022}. A slightly different approach for the data-driven automatic discovery of constitutive laws was presented by Flaschel et al. using sparse regression based on displacement and global force data \cite{Flaschel2021}. This work was further extended by physically consistent deep neural networks for the discovery of isotropic and anisotropic hyperelastic constitutive laws \cite{Thakolkaran2022}.  Besides, Bahmani and Sun presented a framework for a physically constrained symbolic modeling approach for polyconvex incompressible hyperelastic materials \cite{bahmani2023physics}. A probabilistic machine learning approach for data-driven isotropic and anisotropic constitutive models was reported by Fuhg and Bouklas \cite{Fuhg2022}. A first study on the prediction of inelastic effects in cross-linked polymers using neural networks was published by Ghaderi et al. \cite{Ghaderi2020}.
 
In spite of clear advantages as a useful tool for the model discovery, artificial neural networks have certain limitations. First of all, they appear as "black boxes" and are thus hard to interpret. Training such models requires significant computational effort, and the high complexity of the resulting models can make further applications, such as finite element simulations, difficult. 	 An alternative method which can overcome many disadvantages of data-driven approaches is the use of symbolic regression (SR). SR is a relatively new regression method that belongs to the class of interpretable machine learning algorithms. It determines a mathematical expression by searching a solution space. In this search space, a best-fitting expression structure is identified for a given data set \cite{Augusto2000}. Accordingly, an expression optimal in terms of simplicity and accuracy with respect to the data set is determined. The major advantage of this approach is that it identifies an analytical model while reducing the effect of human bias.  Application examples of SR include the modeling of dynamical systems in physics [26], as well as the discovery of governing equations in applied mechanics \cite{huang2021ai} or even the reconstruction of orbital anomalies \cite{manzi2020orbital}. SR has also been used for constitutive modeling of plastic deformations in metals by Kabliman et al. \cite{kabliman2019prediction, kabliman2021application}. Bomarito et al. \cite{bomarito2021development} developed plasticity models using data from micro-mechanical finite element simulations. Wang et al. \cite{wang2022establish} further proposed an approach using a tensorial sparse symbolic regression method. A first approach incorporating theoretical foundations of continuum mechanics for the prediction of interpretable hyperelastic material models was reported by Abdusalamov et al. \cite{Abdusalamov2023}.

Several libraries exist for SR, starting with a symbolic regression framework based on Genetic Programming (GP), describing the evolutionary component of the algorithm \cite{Koza1994}. A detailed review of the advantages and shortcomings of these libraries is presented in \cite{la2021contemporary}. One of the most promising approaches is the deep symbolic optimization (DSO) framework by Petersen et al. \cite{petersen2019deep}, where a novel method combining recurrent neural networks for the prediction of algebraic equations is proposed and a "white box" model is created using a "black box".

In this work, we present a novel approach using deep symbolic regression (DSR) to generate material models describing the Mullins effect in the context of nearly incompressible hyperelastic materials. The two steps framework first identifies a strain energy function for the primary loading of the material. Subsequently, a damage function describing unloading, secondary and further loading responses is identified. The efficiency of the presented approach is demonstrated through benchmark tests with the generalized Mooney-Rivlin model and the Ogden-Roxburgh model. A thorough investigation of the generalizability and robustness of the framework is presented.  The proposed method is capable of rediscovering the underlying strain energy functions provided for multi-axial loading conditions with extremely high $R^2$ values. Moreover, the framework relies only on sparse input data. In addition, the proposed method is extensively validated on a temperature-dependent data set, demonstrating its versatile and reliable performance.

The paper is organized as follows: \autoref{sec:Methodology} discusses the proposed methodology. First, the continuum mechanical framework is discussed and an overview of deep symbolic regression is given. The implementation of the framework is explained in \autoref{sec:implementation}. In \autoref{sec:ResultsDiscussion} the performance of the proposed approach for several benchmarks and a temperature-dependent data set is discussed. Finally, a brief conclusion highlighting the main aspects of this work is presented in \autoref{sec:Conclusion}.

\section{Methodology}
\label{sec:Methodology}

\subsection{Continuum Mechanical Framework}
\label{sec:CM}
In the following, we consider a strain energy function $\Psi(\IC, \IIC, \IIIC)$ of an isotropic hyperelastic material expressed in terms of the principal invariants $\IC, \IIC$ and $\IIIC$ of the right Cauchy-Green tensor $\tens{C}=\tens{F}^\text{T}\tens{F}$, where $\tens{F}$ denotes the deformation gradient. Beyond the hyperelasticity the strain energy can depend additionally on some other variables such as temperature,
strain rate, damage or loading history parameters. The first Piola-Kirchhoff stress tensor $\tens{P}_0$ can be represented by
\begin{equation}
	\label{equ:S}
	\tens{P}_0
	= 2 \tens{F}\frac{\partial \Psi(\tens{C})}{\partial \tens{C}} 
	=2\left[\left(\frac{\partial \Psi}{\partial \IC} + \IC \frac{\partial \Psi}{\partial \IIC}\right) \mathbf{F}-\frac{\partial \Psi}{\partial \IIC} \tens{FC} + \IIIC \frac{\partial \Psi}{\partial \IIIC} \tens{F}^{-\text{T}}\right].
\end{equation}
The function $\Psi(\tens{C})$  should satisfy the conditions of the energy and stress free natural state at $\mathbf{F}=\mathbf{I}$. Accordingly,
\begin{equation}\label{stress-free}
	\Psi(\mathbf{I})=0, \quad  
	\left. \frac{\partial \Psi(\tens{C})}{\partial \tens{C}}\right|_{\tens{C}=\tens{I}}=\tens{0}.
\end{equation}

In the case of nearly incompressible behavior it is reasonable to decompose the deformation gradient multiplicatively into a volumetric $J\tens{I}$ and an isochoric part $\tens{\bar F}=J^{-1/3}\tens{F}$ according to \cite{Flory1961}, where  $J= \det\tens{F}$. 
The principal invariants of the isochoric right Cauchy-Green tensor $\tens{\bar C}=\tens{\bar F}^{\text{T}}\tens{\bar F}$ take the form $\ICdev= J^{-2/3}\mathrm{I}_{\tens{ C}}$, $\IICdev = J^{-4/3}\mathrm{II}_{\tens{ C}}$ and $\IIICdev = 1$. Accordingly, the first Piola-Kirchhoff stress tensor can be expressed by 

\begin{align} \label{P-v-i-split}
	\tens{P}_0  = & 2 \left(\frac{\partial \Psi}{\partial \ICdev}  + \ICdev \frac{\partial \Psi}{\partial \IICdev}\right) J^{-2/3} \mathbf{F} - 2 \frac{\partial \Psi}{\partial \IICdev} J^{-4/3}  \tens{FC}  + J \left(\frac{\partial \Psi}{\partial J}-\frac{2}{3J} \frac{\partial \Psi}{\partial \ICdev}  \ICdev   - \frac{4}{3J} \frac{\partial \Psi}{\partial \IICdev} \IICdev\right) \tens{F}^{-\text{T}}.
\end{align}
Assuming that the volumetric and isochoric responses are independent of each other, the strain energy function can further be decomposed by $\Psi(\tens{C}) = \bar \Psi (\mathrm{I}_{\tens{\bar C}}, \mathrm{II}_{\tens{\bar C}})+ \hat \Psi(J)$. For the volumetric strain energy $\hat \Psi(J)$ many reliable formulations satisfying \eqref{stress-free} and
convexity conditions have been proposed (see e.g. \cite{HARTMANN20032767}).

For incompressible materials characterized by the constraint $J=1$ the constitutive equation takes the form
\begin{align}\label{P-incompr}
	\mathbf{P}_0=2 \tens{F}\frac{\partial \Psi(\tens{C})}{\partial \tens{C}} -p \tens{F}^{-\text{T}}
	=2\left[\left(\frac{\partial \Psi}{\partial \IC} + \IC \frac{\partial \Psi}{\partial \IIC}\right) \mathbf{F}-\frac{\partial \Psi}{\partial \IIC} \tens{FC} \right]
	-p \tens{F}^{-\text{T}}, 
\end{align}
where an additional hydrostatic pressure $p$ can be determined from equilibrium and boundary conditions. 

Expressions \eqref{equ:S}, \eqref{P-v-i-split} or \eqref{P-incompr} are used for fitting to 
experimental or artificially created strain-stress data. Therein, only the derivatives of the strain energy density function with respect to the invariants are material-specific and will be used to identify an optimal expression of the strain energy function.

In the above mentioned pseudo-elastic model of the Mullins effect by Ogden and Roxburgh \cite{OgdenRoxburgh1999} the stress resulting from the hyperelastic constitutive equation is
reduced by a damage variable $\eta$ as follows
\begin{equation}
	\label{equ:S0}
	\tens{P} = \eta \left(\Psi_{\text{max}}, \Psi \right) \tens{P}_{0}.
\end{equation}
Accordingly, $\eta$ depends on the actual $\Psi$ and the maximal value $\Psi_{\text{max}}$ of the strain energy previously reached in the loading history and is approximated by 
\begin{equation}
	\label{equ:Eta}
	\eta \left(\Psi_{\text{max}}, \Psi \right) = 1 - \frac{1}{r} \erf \left( \frac{\Psi_{\text{max}} -  \Psi  }{m + \beta \Psi_{\text{max}}} \right),
\end{equation}
where $\erf(\cdot)$ represents the error function while $r$, $\beta $ and $m$ denote material constants.
 
\subsection{Deep Symbolic Regression}
\label{sec:DSR}
\begin{figure}[ht!]
	\centering
	\includegraphics[width=0.75\textwidth]{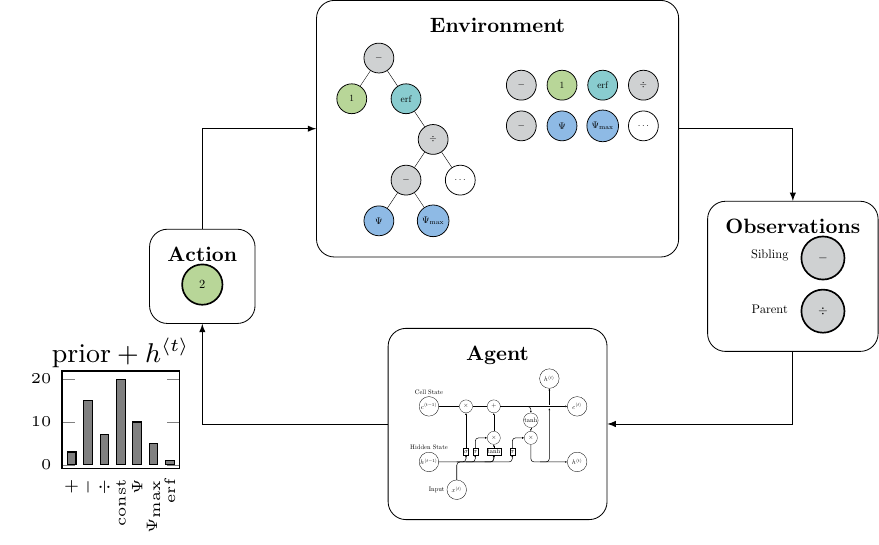}
	\caption{Visualization of the deep symbolic regression process. The environment consists of a traversal of tokens where the last entry is sampled through the RNN environment. The neural network receives as input observations of the sibling and parent of the current token. The output of the agent is a probability distribution function that is used to sample the next action.}
	\label{fig:DSOProcess}
\end{figure}

Peterson et al. \cite{Petersen2021} reported a novel method of using a recurrent neural network (RNN) to predict a mathematical expression based on a sampled distribution through a risk-seeking policy gradient. In general, a mathematical expression can be represented as a graphical tree where internal nodes are mathematical operators or functions and terminal nodes are input variables or constants. In the case of deep symbolic regression, each expression tree is transformed into a sequence of node values called "tokens" using its pre-order traversal, as for example visualized for a strain energy function in \autoref{fig:DSOProcess}. The nodes of the traversal correspond to either operations, functions, constants, or arguments. The recurrent neural network is trained on a hierarchical input containing information about the entire expression tree. Therefore, observations about siblings and parents are used to train the RNN. The next element of the traversal is sampled based on a probability distribution function (see \autoref{fig:DSOProcess}). Moreover, several constraints are included to limit the search space to a manageable size. These are in particular limits of the maximal depth of the expression tree, of the set of allowed operators and functions, of the set of allowed input variables and constants. Besides, a prior in the sampled probability distribution is included. The package also provides a constant optimization option. Once a prior is sampled, the corresponding symbolic expression is instantiated and evaluated. The fitness measure is estimated based on the normalized root mean square error (NRMSE) and used as a reward signal for the next sampling episode.

\section{Implementation}
\label{sec:implementation}
Here, we are going to combine the continuum mechanical framework briefly presented in \autoref{sec:CM} with the concepts of deep symbolic regression as discussed in \autoref{sec:DSR}. To this end, several implementation aspects need to be addressed, such as the evaluation of the loss function, the differentiation of the obtained strain energy functions, etc. 

\begin{figure}[h!]
	\centering
	\includegraphics[width=0.8\textwidth]{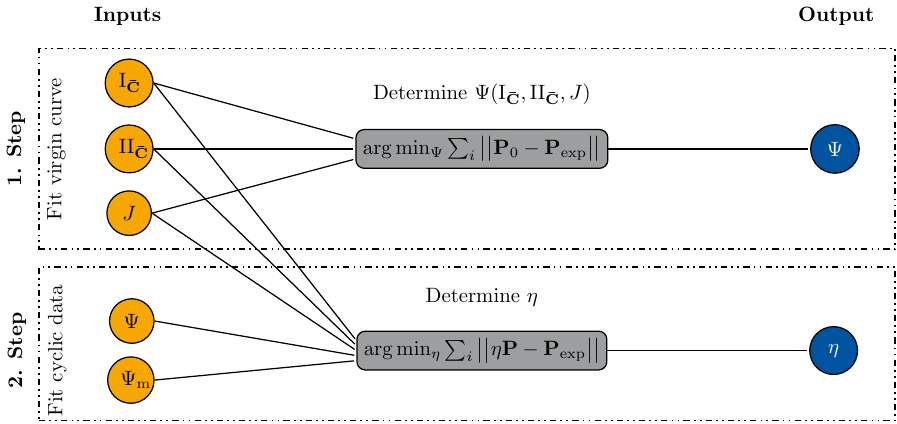}
	\caption{Visualization of the implemented two step procedure. In the first step, the strain energy function $\Psi$ is expressed in terms of the invariants and other parameters using the data of the primary loading curve. In the second step, the damage function $\eta$ is determined using all inputs from the first step as well as the values of $\Psi$ and $\Psi_m$. The fit is performed on the cyclic loading data.}
	\label{fig:MullinsProcess}
\end{figure}

For the sake of flexibility, the modeling process is split into two separate steps: first the hyperelastic material model is determined.
Both the isochoric and volumetric contributions to the strain energy function can separately be determined. These contributions serve as the input in the second step where the damage function is searched and evaluated, see \autoref{fig:MullinsProcess}. 

The strain energy is determined on the basis of stress-strain data of the primary loading (the virgin curve). To this end, a user-defined loss function calculating the derivatives of the strain energy function with respect to the given invariants is defined. The differentiation is performed numerically using the finite difference scheme. For comparison and validation purposes we also implemented the reverse-mode algorithmic differentiation by incorporating it into the call tree. However, it appeared $60$\% less numerically effective in comparison to the numerical differentiation without any noticeable improvement in the quality of the output. The framework provided in \cite{Abdusalamov2023} was further used to compute the general responses for different loading cases and material models. Moreover, the process was accelerated by vectorizing several operations using \textsc{NumPy} broadcasting operations. The loss function is evaluated based on the provided stress response and stresses calculated by \eqref{equ:S}, \eqref{P-v-i-split} or \eqref{P-incompr}. In this sense, the first Piola-Kirchhoff stress tensor expressed in these formulas appears to be very convenient since it allows a direct comparison with the experimental stress response.

Since for this comparison with experimental or artificially created stress data only derivatives of the strain energy functions are relevant, the condition \eqref{stress-free}$_1$ can easily be satisfied by correction of the resulting expression by a constant. \eqref{stress-free}$_2$ is fulfilled automatically by including the point $\tens{P}=\tens{0}$ at $\tens{F}=\tens{I}$ into the set of the data used for the search of the mathematical expression of the strain energy function. 

The provided basis functions was given by ["\textsf{add}", "\textsf{sub}",  "\textsf{n2}", "\textsf{mul}", "\textsf{div}", "\textsf{sqrt}", "\textsf{exp}", "\textsf{log}", "\textsf{const}"] and ["\textsf{add}", "\textsf{sub}", "\textsf{mul}", "\textsf{tanh}", "\textsf{div}", "\textsf{erf}", "\textsf{const}"] in the first and second step, respectively. Note that the last term in both steps executes the constant optimization. Despite the increased time requirement and the over-fitting risk, it allows a significantly higher rate of expression recovery. All hyperparameters have been left at their default settings, except for the total number of samples, which has been set to a maximum of $\SI{75000 }{}$. For a more reliable performance, the final evaluation step contains a filter which rounds all decimal numbers to the second digit leading to a simplification of determined expressions. Introducing more invariants (as for example in the case of anisotropic material response) or additional physical parameters (such as temperature) appears fairly straightforward, since only the input space needs to be extended correspondingly in the first step.
\section{Results and Discussion}
\label{sec:ResultsDiscussion}

\subsection{Benchmark Test with the Generalized Incompressible Mooney-Rivlin Model}
\label{subsec:gMR}
The generalized Mooney-Rivlin model \cite{Bower2010,Mooney1940} covers
a large variety of rubber-like materials and can thus be used as benchmark test to evaluate the performance of the proposed framework in determining the expression of the strain energy function. It is given for this model by
\begin{equation}
	\label{equ:gmR}
	\Psi_{\rm gMR} = 
	\sum\limits_{i=1}^3 \left[c_{i0}\left(\IC-3\right)^i + c_{0i}\left(\IIC-3\right)^i \right],
\end{equation}
where $c_{i0}$ and $c_{0i}$ represent material constants. In choosing their values, we considered three cases of increasing model complexity. These are given in \autoref{tab:gMR_par}.
\begin{table}
	\centering
	\caption{Sets of material constants for the generalized Mooney-Rivlin model with increasing complexity.}
	\label{tab:gMR_par}
	\begin{tabular}{ccccccc}
		\toprule
		Case & $c_{10}$ & $c_{20}$ & $c_{30}$ & $c_{01}$ & $c_{02}$ & $c_{03}$ \\
		& $ \left[ \SI{}{\mega \Pa} \right] $ & $ \left[ \SI{}{\mega \Pa} \right] $ & $ \left[ \SI{}{\mega \Pa} \right] $ & $ \left[ \SI{}{\mega \Pa} \right] $ & $ \left[ \SI{}{\mega \Pa} \right] $ & $ \left[ \SI{}{\mega \Pa} \right] $   \\
		\midrule
		1 & 0.63 & 0.00 & 0.00 & 0.39 & 0.00 & 0.00  \\ 
		2 & 0.95 & 0.66 & 0.00 &  0.51 & 0.62 & 0.00  \\ 
		3 & 0.73  & 0.43 & 0.1 & 0.99 & 0.97 & 0.32  \\
		\bottomrule
	\end{tabular}
\end{table} 

\begin{figure}[ht!]
	\centering
	\begin{subfigure}{0.45\textwidth}
		{\includegraphics[width=0.97\textwidth]{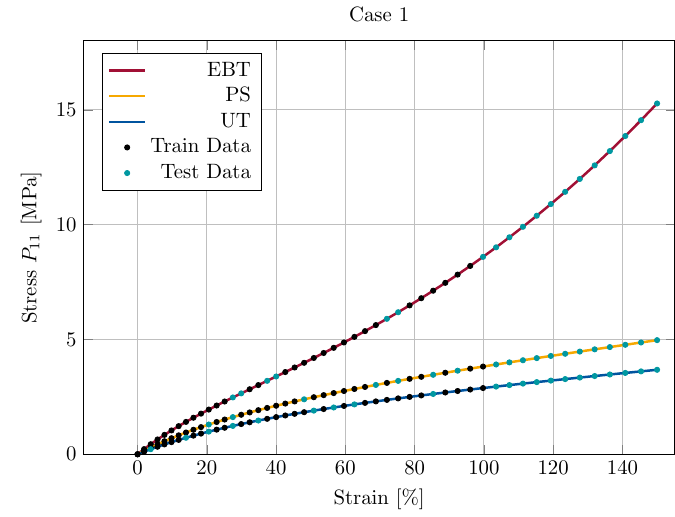}}
		\caption{}
		\label{fig:MR2IncompStress}
	\end{subfigure}
	\quad
	\begin{subfigure}{0.45\textwidth}
		\centering
		\includegraphics[width=1\textwidth]{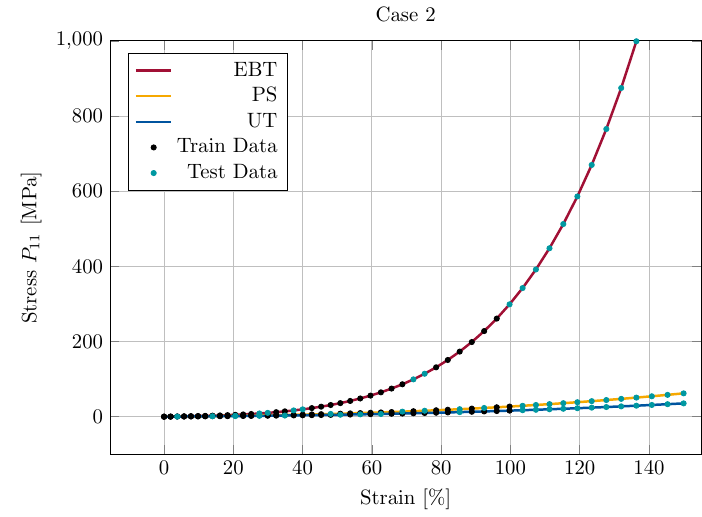}
		\caption{}
		\label{fig:MR4IncompStress}
	\end{subfigure}
	\hfill
	\begin{subfigure}{0.45\textwidth}
		\centering
		\includegraphics[width=1\textwidth]{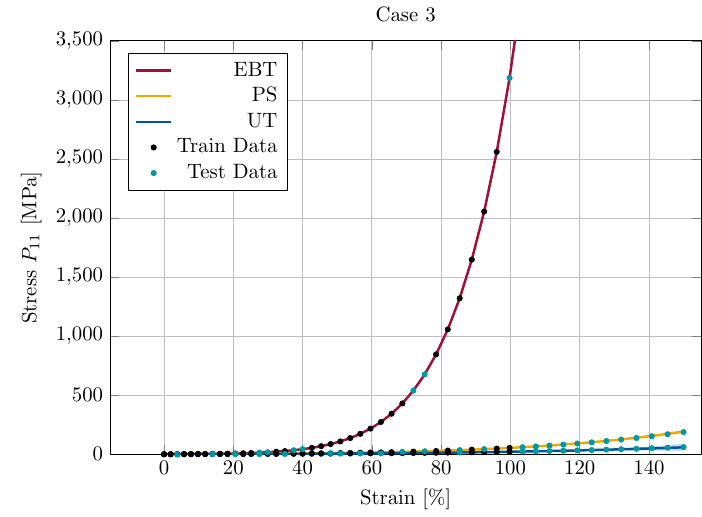}
		\caption{}
		\label{fig:MR6IncompStress}
	\end{subfigure}
	\caption{Comparison of the mean stress-strain response of five models (\autoref{tab:gMRPredictions}) obtained by DSR with the corresponding training and test data for UT, PS and EBT based on the generalized Mooney-Rivlin model for all three sets of material constants. Color surroundings of the curves reflect $6\sigma$ confidence intervals of the predictions.}
	\label{fig:GenMonRivlinCases}
\end{figure}

For every of these sets of material constants, the responses are determined under uniaxial (UT), equibiaxial tension (EBT) and pure shear (PS). For each of these loading cases only 50 data points we calculated. In order to evaluate the interpolation quality of resulting expressions of the strain energy function only the data in the interval of strains between 0 and ${100}$\% were used for fitting with an additional train ($80$\%) and test ($20$\%) split of the data. The data in the strain interval between $100$\% and $150$\% were used to evaluate the extrapolation capability of the resulting expressions.
The performance of the models in the three cases can be observed in \autoref{fig:MR2IncompStress}, \autoref{fig:MR4IncompStress} and \autoref{fig:MR6IncompStress}, where the resulting stress-strain diagrams are plotted against the train and test data. For each set of the material constants five expressions of the strain energy function
given in \autoref{tab:gMRPredictions} were determined. The  predictions visualized in \autoref{fig:MR2IncompStress}, \autoref{fig:MR4IncompStress} and \autoref{fig:MR6IncompStress} correspond to the mean response of all five resulting models. Furthermore, the $6\sigma$ confidence intervals are also shown in these diagrams as color surroundings of the curves.
An interesting insight is that for the first two and the last case the recovery rate of the underlying strain energies is of $100$\% and $0$\%, respectively, although all models have a $R^2$-score higher than $99.98$\%. This $0$\% recover rate in combination with almost $100$\% of $R^2$-score suggest that DSR converges to an alternative expression of the strain energy function which is possible due to high nonlinearity of the problem in particular in the last case of the generalized Mooney-Rivlin model. For a discussion of such nonuniqueness for example in context of the Ogden model we refer to \cite{ogden2004fitting}. Nevertheless, close approximations can be found to describe the data sets created by these models with high precision. Note also that the extrapolation quality of the underlying models is excellent although very few training data points have been used. In comparison to the previous work \cite{Abdusalamov2023}, the confidence interval is everywhere negligible and is only visible under UT in the last case, which is also reflected by the high $R^2$-scores. Thus, the proposed framework demonstrates robust performance by extrapolating and interpolating the sparse data provided for all three complexity cases and three different loading conditions.   

\subsection{Benchmark Test for Nearly Incompressible Mooney-Rivlin Model}
The above described volumetric-isochoric split of the strain energy function is a very important tool especially for finite element software and deserves thus a separate numerical example. In this example, the isochoric part of the strain energy is again given by the Mooney-Rivlin model \eqref{equ:gmR} with the same constants as in the first case in \autoref{tab:gMR_par}. For the volumetric response, we apply a classical quadratic function in $J$ and an expression by Miehe \cite{miehe1994aspects} as follows
\begin{align}\label{vol-energies}
\hat{\Psi}_{1} = \frac{1}{2} \kappa_{1}  (J - 1)^2 \quad \text{and} \quad  \hat{\Psi}_{2} = \kappa_{2}\left(J - \ln J - 1\right).
\end{align}
The constants $\kappa_{1}=50$ and $\kappa_{2}=65$ are set in such a way that the equibiaxial tension response of both volumetric energies is identical at $\SI{150}{\percent}$ of strain.

In both cases, values of $R^2$ were higher than $99.70$\% for all five models (\autoref{tab:gMRPredictionsComp}) obtained by DSR. Interestingly, the recovery rate of the volumetric functions is $60$\% for the first formulation and $0$\% for the second one. Similarly to the case of the isochoric strain energy function, this indicates nonuniqueness of the volumetric contribution accurately describing stress-strain data. In the applied  procedure both the isochoric and volumetric strain energies were simultaneously searched for. A reasonable alternative would thus be to directly specify a numerically stable volumetric strain energy and identify only  the isochoric contribution. Note also that for the first formulation the isochoric parts are correctly identified for 4 out of 5 samples. For the second case, the isochoric parts are not fully recovered.

\begin{figure}[ht!]
	\centering
	\begin{subfigure}{0.45\textwidth}
		{\includegraphics[width=1\textwidth]{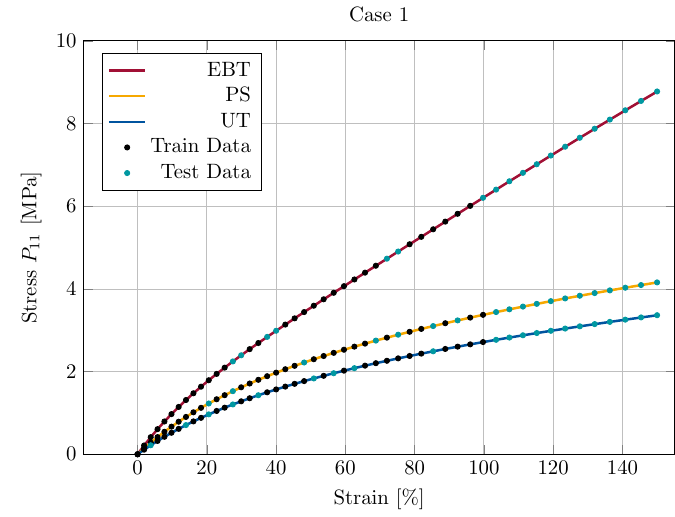}}
		\caption{}
		\label{fig:MR2CompStress1}
	\end{subfigure}
	\quad
	\begin{subfigure}{0.45\textwidth}
		\centering
		\includegraphics[width=1\textwidth]{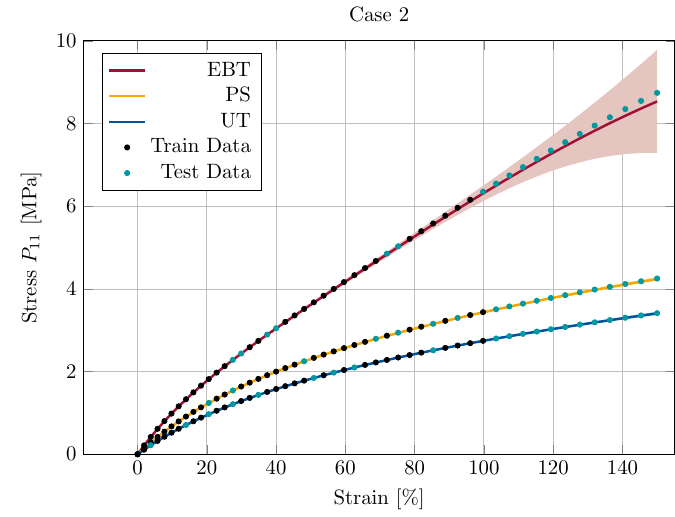}
		\caption{}
		\label{fig:MR2CompStress2}
	\end{subfigure}
	\caption{Comparison of the mean stress-strain response of five models obtained by DSR with the corresponding training and test data for UT, PS and EBT for the two different volumetric strain energy functions \eqref{vol-energies}. Color surroundings of the curves reflect $6\sigma$ confidence intervals of the predictions.}
	\label{fig:GenMonRivlinComp}
\end{figure}

\subsection{Benchmark Test with the Ogden-Roxburgh Model}
\begin{figure}[ht!]
	\centering
	\begin{subfigure}{0.45\textwidth}
		{\includegraphics[width=1\textwidth]{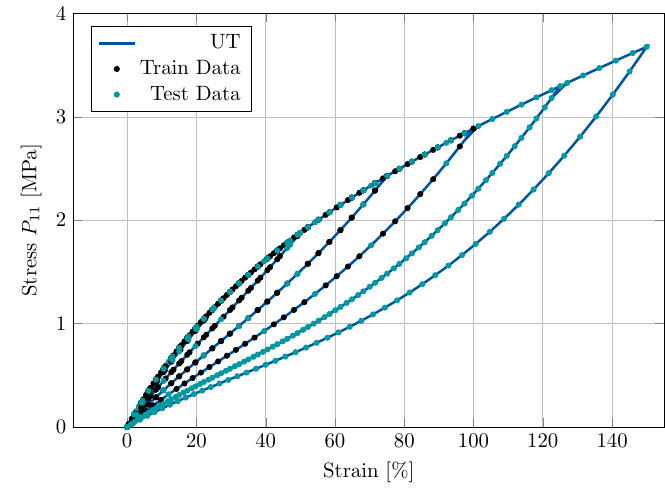}}
		\caption{}
		\label{fig:MullinsUT}
	\end{subfigure}
\quad
	\begin{subfigure}{0.45\textwidth}
		\centering
		\includegraphics[width=1\textwidth]{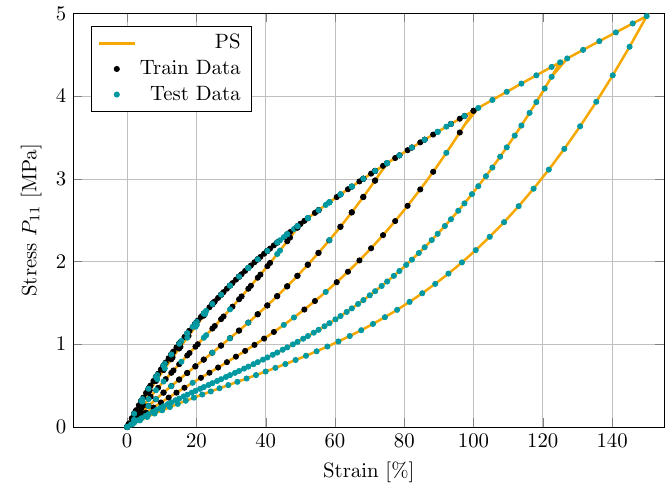}
		\caption{}
		\label{fig:MullinsPS}
	\end{subfigure}
\hfill
	\begin{subfigure}{0.45\textwidth}
		\centering
		\includegraphics[width=1\textwidth]{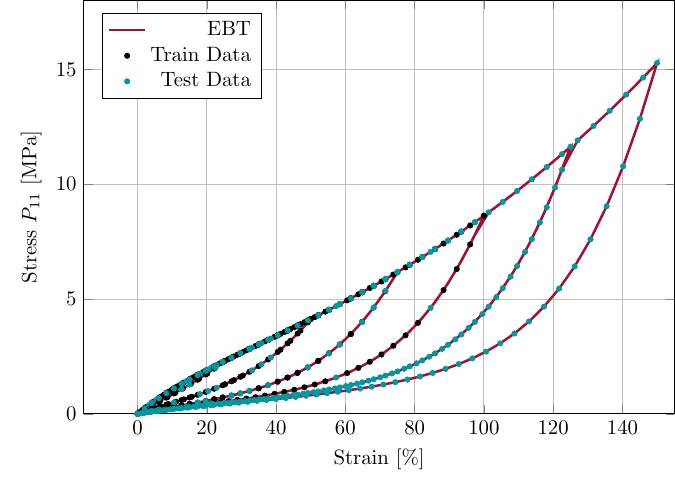}
		\caption{}
		\label{fig:MullinsEBT}
	\end{subfigure}
	\caption{Comparison of the mean stress-strain response of five models (\autoref{tab:OR}) obtained by DSR with the
		corresponding training and test data for UT, PS and EBT based on the generalized Mooney-Rivlin of case 1 for the primary loading and the Ogden-Roxburgh model for the softening (unloading and further reloading.) Color surroundings of the curves reflect $6\sigma$ confidence intervals of the predictions.}
	\label{fig:OR_Fit}
\end{figure}

In order to evaluate the performance of the proposed framework in describing softening behavior of elastomers we applied 
the Ogden-Roxburgh model briefly presented above. As described in \autoref{sec:implementation} (see also \autoref{fig:MullinsProcess}) the modeling process starts with evaluation of the strain energy functions based on the primary loading. Stress-strain data for this loading were created by the generalized incompressible Mooney-Rivlin model
with the material constants for the case 1 (\autoref{tab:gMR_par}). Stress-strain data for unloading and further loading were generated by Ogden-Roxburgh model \autoref{equ:Eta} with the following material constants $r={1.2}$, $\beta={0.5}$ and $m=2$. We applied cyclic loading with step-wise increasing amplitude as shown in \autoref{fig:OR_Fit}. In these diagrams stresses as well as the corresponding $6\sigma$ confidence intervals are plotted against strains for the loading cases of UT, PS and EBT. For each loading case a test split of ${20}$\% was used in the strain domain between 0 and ${100}$\% to access the interpolation quality. The additional test range between ${100}$\% to ${150}$\% serves to evaluate the quality of extrapolation. Accordingly, only first three unloading curves (from the strain amplitude up $100$\%) were used to identify the damage function $\eta$ in \eqref{equ:S0}. The remaining two loading cycles (with the strain amplitude over $100$\%) are used to evaluate the quality of the fit.

The recovery rates and the $R^2$ score for the virgin curve are identical to the  results obtained in \autoref{subsec:gMR}. For the damage variable an average of $R^2 = {99.98}$\% is achieved. The recovery rate of $80$\% was reached. The so determined damage models are listed in \autoref{tab:OR}. Based on this benchmark test we can be conclude that only few loading cycles with sparse data are sufficient to recover the softening behavior with the extremely high accuracy. The underlying continuum mechanical framework is able both to interpolate and extrapolate the given data and determine the damage function in a robust manner.

\subsection{Application to Experimental Temperature Dependent Data Set}
\begin{figure}[ht!]
	\centering
	\begin{subfigure}{0.45\textwidth}
		{\includegraphics[width=1\textwidth]{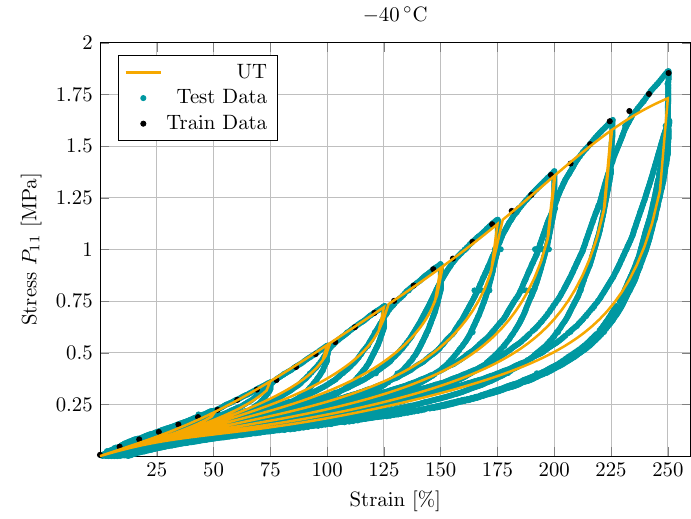}}
		\caption{}
		\label{fig:Temp_1}
	\end{subfigure}
	\hfill
	\begin{subfigure}{0.45\textwidth}
		\centering
		\includegraphics[width=1\textwidth]{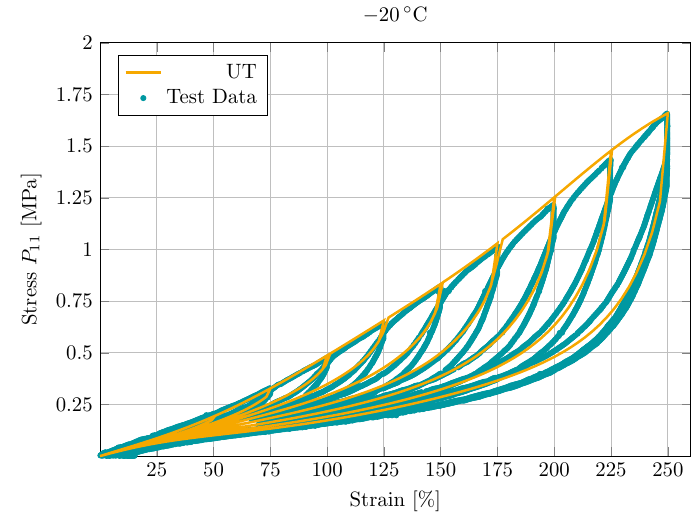}
		\caption{}
		\label{fig:Temp_2}
	\end{subfigure}
	\begin{subfigure}{0.45\textwidth}
		{\includegraphics[width=1\textwidth]{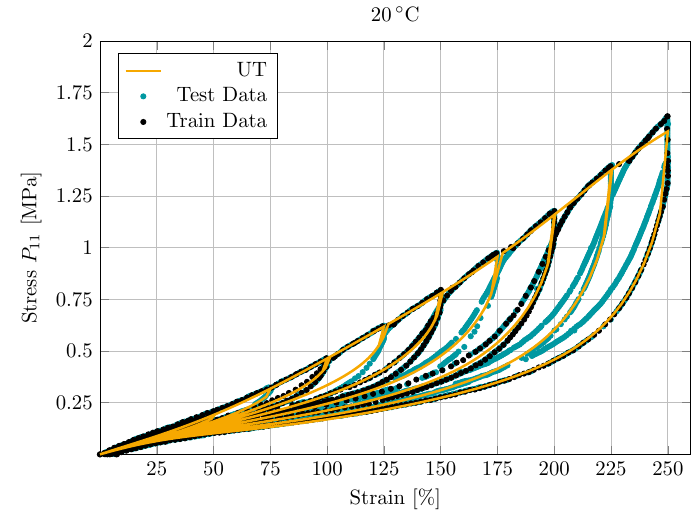}}
		\caption{}
		\label{fig:Temp_3}
	\end{subfigure}
	\hfill
	\begin{subfigure}{0.45\textwidth}
		\centering
		\includegraphics[width=1\textwidth]{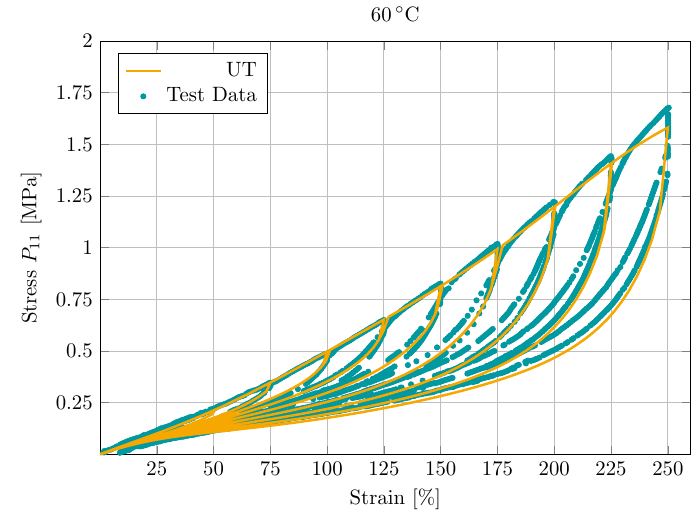}
		\caption{}
		\label{fig:Temp_4}
	\end{subfigure}
	\begin{subfigure}{0.45\textwidth}
		{\includegraphics[width=1\textwidth]{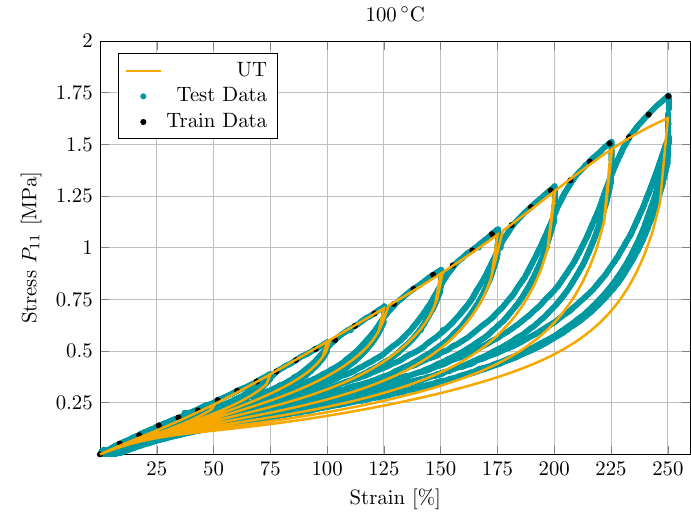}}
		\caption{}
		\label{fig:Temp_5}
	\end{subfigure}
	\hfill
	\begin{subfigure}{0.45\textwidth}
		\centering
		\includegraphics[width=1\textwidth]{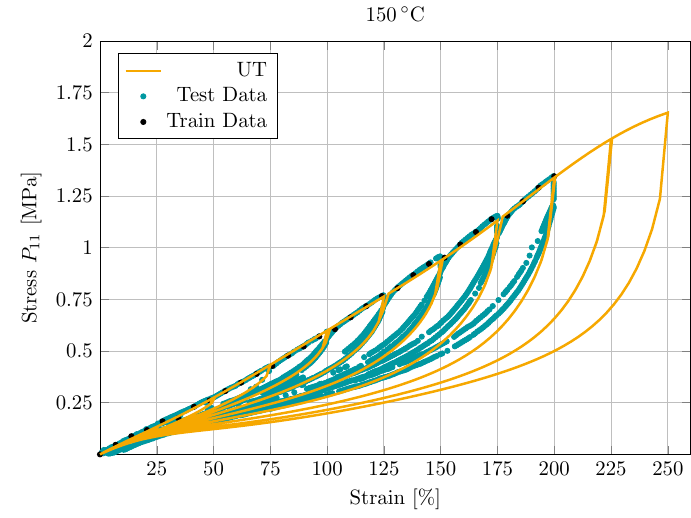}
		\caption{}
		\label{fig:Temp_6}
	\end{subfigure}
	\caption{Cyclic uniaxial tension of filled silicone with stepwise increasing strain amplitude at different temperatures: comparison of the stress-strain responses resulting from the model \eqref{SEF-Temp}, \eqref{eta-Temp} obtained by DSR with the experimental data \cite{rey2013influence}. Primary loading curves for the temperatures $-40^\circ$C, $20^\circ$C, $100^\circ$C and $150^\circ$C as well as one loading cycle for the temperature $20^\circ$C were used as training data.}
	\label{fig:Temp_fit}
\end{figure}

Finally, we consider an experimental temperature dependent data set \cite{rey2013influence}. In these experiments filled silicone specimens were subject to uniaxial tension under temperatures $-40^\circ$C, $-20^\circ$C, $20^\circ$C, $60^\circ$C, $100^\circ$C and $150^\circ$C. Loading-unloading cycles with step-wise increasing amplitude were applied. The data set consists of only 120 data points and was used as follows. The temperature dependent strain energy function was determined from the primary loading curves under the temperatures $-40^\circ$C, $20^\circ$C, $100^\circ$C and $150^\circ$C, while the curves of $-20^\circ$C and $60^\circ$C served as test data. The material was considered as ideally incompressible. Accordingly, we obtained
\begin{equation}
	\begin{split}
		\Psi_{\text{T}} (\mathrm{I}_{\tens{C}}, \mathrm{II}_{\tens{C}}, \tilde{T}) =	0.11 \sqrt[4]{\mathrm{I}_{\tens{C}}} \biggl( \mathrm{I}_{\tens{C}} - 0.24 &\sqrt{0.85 \biggl( \frac{\mathrm{II}_{\tens{C}}}{\tilde{T}} + \biggl(2.71\frac{\mathrm{I}_{\tens{C}} \mathrm{II}_{\tens{C}}}{\tilde{T}} +  \log{\left(0.84\frac{\mathrm{I}_{\tens{C}} \mathrm{II}_{\tens{C}}}{\tilde{T}} + \log{\tilde{T}} \right)}\biggr) \biggr)} \\ &\cdot \sqrt{\exp \left(0.55\mathrm{I}_{\tens{C}} \left(- 2.28 \tilde{T} + \log{\tilde{T}} + 1.64\right)\right) - 1.03} - 2.65 \biggr),
	\end{split}
	\label{SEF-Temp}
\end{equation}
where $\tilde{T} = \nicefrac{T}{300} + \nicefrac{1}{3}$ is a scaled temperature mapping the temperature range from $-100^\circ$ to $200^\circ$ to the data range between $0$ and $1$. Although rescaling the temperature range to a comparable order of magnitude is not mathematically necessary, it improves the computational efficiency of the optimization process. As explained in \autoref{sec:implementation} the obtained strain energy function \eqref{SEF-Temp} is shifted such that the condition \eqref{stress-free}$_1$ is satisfied. The shift is expressed as 
\begin{align}
	\Psi_{\text{T}}^{n} (\mathrm{I}_{\tens{C}}, \mathrm{II}_{\tens{C}}, \tilde{T}) = \Psi_{\text{T}} (\mathrm{I}_{\tens{C}}, \mathrm{II}_{\tens{C}}, \tilde{T}) - \Psi_{\text{T}} (3, 3, \tilde{T}), 
\end{align}
where the new function $\bar\Psi_{\text{T}}^{n} (\mathrm{I}_{\tens{C}}, \mathrm{II}_{\tens{C}}, \tilde{T})$ is used for identifying the temperature dependent damage function of the Ogden-Roxburg model \eqref{equ:S0}. To this end, only unloading and reloading stress-strain data of the cyclic test for the temperature $20^\circ$C were used. All remaining cyclic data sets in particular for the temperatures $-20^\circ$C $60^\circ$C served as test data. The so obtained damage function is of the form
\begin{equation}\label{eta-Temp}
	\begin{split}
\eta_{T} \left(\Psi_{\text{T}}^{n}, \Psi_{\text{T,max}}^{n}\right) =	
&0.29 \coth\biggl[\Psi_{\text{T}}^{n}- 0.37 \Psi_{\text{T}}^{n}\biggl(\frac{\Psi_{\text{T,max}}^{n}}{0.09 + 0.33 \Psi_{\text{T}}^{n}} + \erf\bigl(1936.97 + \Psi_{\text{T,max}}^{n}\bigr) - \tanh\bigl(\Psi_{\text{T}}^{n}\bigr)\biggr) \\ &+ \tanh\bigl(1.33 - \tanh\big(1.90 - 14.49 \Delta\big) - \tanh\left[\tanh\left(1.67 - 47186.20 \Delta\right)\right]\bigr)\biggr],
\end{split}
\end{equation}
where $\tanh(\cdot)$ and $\coth(\cdot)$ denote hyperbolic tangent and cotangent, respectively,   $\Psi_{\text{T,max}}^{n}$ is the maximal value of the accumulated strain energy while $\Delta = \erf\bigl(\Psi_{\text{T}}^{n}- \Psi_{\text{T,max}}^{n}\bigr)$. The stress-strain responses resulting from \eqref{SEF-Temp}, \eqref{eta-Temp} according to \eqref{P-incompr} and \eqref{equ:S0} are shown in \autoref{fig:Temp_fit} for all temperatures. It is seen that
only the primary curve and only one loading cycle were sufficient to describe very accurately the stress-strain responses at all six temperatures.

\section{Conclusion}
\label{sec:Conclusion}

In this work, we presented a novel approach based on deep symbolic regression to generate material models describing the softening behavior and in particular the Mullins effect in elastomers. Several benchmark tests both with ideally and nearly incompressible materials confirm the ability of the proposed framework to determine the strain energy and damage function accurately describing both experimental and artificially created stress-strain data. In spite of a low recovery rate in the case of a hyperelastic model with the highest complexity, the $R^2$ values indicate the efficiency of the procedure in determining accurate model approximations. This concerns also the description of the inelastic behavior. Indeed, the proposed framework excelled in very accurate recovering of cyclic stress-strain curves created by the Ogden-Roxburgh model and maintained robust performance even with sparse temperature-dependent experimental data. The stress-strain responses for different temperatures were well described, indicating excellent performance of the framework. 

Overall, our approach demonstrates high potential of deep symbolic regression combined with continuum mechanical framework in both elastic and inelastic material modeling even when limited amount of data are available.




%
%



\appendix
	\begin{table}
	\caption{Predicted strain energies for benchmark test using generalized Mooney-Rivlin model for all three cases.}
	\label{tab:gMRPredictions}
	\begin{tikzpicture}
		\node[rotate=90,transform shape, color=black]{
			\renewcommand{\arraystretch}{2}
			\begin{tabular}{cccc}
			\toprule
			Case & N & Strain energy & $R^2$-score \\
			\midrule
			1 & 1 & $\bar\Psi_{\rm gMR} = 0.63\mathrm{I}_{\tens{\bar C}} + 0.39\mathrm{II}_{\tens{\bar C}} + 0.39$ & 1.00 \\
			1 & 2 & $\bar\Psi_{\rm gMR} = 0.63\mathrm{I}_{\tens{\bar C}} + 0.39\mathrm{II}_{\tens{\bar C}} + 0.83$ & 1.00 \\
			1 & 3 & $\bar\Psi_{\rm gMR} = 0.63\mathrm{I}_{\tens{\bar C}} + 0.39\mathrm{II}_{\tens{\bar C}} + 0.61$ & 1.00 \\
			1 & 4 & $\bar\Psi_{\rm gMR} = 0.63\mathrm{I}_{\tens{\bar C}} + 0.39\mathrm{II}_{\tens{\bar C}} - 0.02$ & 1.00 \\
			1 & 5 & $\bar\Psi_{\rm gMR} = 0.63\mathrm{I}_{\tens{\bar C}} + 0.39\mathrm{II}_{\tens{\bar C}} - 2.00$ & 1.00 \\
			\midrule
			2 & 1 & $\bar\Psi_{\rm gMR} = 0.66\mathrm{I}_{\tens{\bar C}}^2 - 3.01\mathrm{I}_{\tens{\bar C}} + 0.62\mathrm{II}_{\tens{\bar C}}^2 - 3.21\mathrm{II}_{\tens{\bar C}} + 1$ & 1.00 \\
			2 & 2 & $\bar\Psi_{\rm gMR} = 0.66\mathrm{I}_{\tens{\bar C}}^2 - 3.01\mathrm{I}_{\tens{\bar C}} + 0.62\mathrm{II}_{\tens{\bar C}}^2 - 3.21\mathrm{II}_{\tens{\bar C}} + 1.0$ & 1.00 \\
			2 & 3 & $\bar\Psi_{\rm gMR} = -(3.0 - 0.66\mathrm{I}_{\tens{\bar C}})(\mathrm{I}_{\tens{\bar C}} - 0.02) + (\mathrm{II}_{\tens{\bar C}}(0.62\mathrm{II}_{\tens{\bar C}} - 3.21) - 1.0)$ & 1.00 \\
			2 & 4 & $\bar\Psi_{\rm gMR} = 0.37\mathrm{I}_{\tens{\bar C}}(1.79\mathrm{I}_{\tens{\bar C}} - 8.15) + 0.62\mathrm{II}_{\tens{\bar C}}(\mathrm{II}_{\tens{\bar C}} - 5.18)$ & 1.00 \\
			2 & 5 & $\bar\Psi_{\rm gMR} = 0.66\mathrm{I}_{\tens{\bar C}}^2 - 3.01\mathrm{I}_{\tens{\bar C}} + \mathrm{II}_{\tens{\bar C}}(0.62\mathrm{II}_{\tens{\bar C}} - 3.21) - 1.0$ & 1.00 \\
			\midrule
			3 & 1 & $\bar\Psi_{\rm gMR} = \frac{1.29  \left(0.25 \mathrm{II}_{\tens{\bar C}} - 0.22\right) \left(\mathrm{I}_{\tens{\bar C}}  \left(0.09 \mathrm{I}_{\tens{\bar C}}^{2} \log{\left(\mathrm{I}_{\tens{\bar C}} \right)} + 0.07 \mathrm{II}_{\tens{\bar C}}\right) + \mathrm{II}_{\tens{\bar C}} \left(\mathrm{II}_{\tens{\bar C}} - 2.67\right) \left(\mathrm{II}_{\tens{\bar C}} - 2.43\right)\right)}{\mathrm{II}_{\tens{\bar C}}} $ & 1.00 \\
			3 & 2 & $\bar\Psi_{\rm gMR} =  \frac{0.3 \mathrm{I}_{\tens{\bar C}} \left(\mathrm{II}_{\tens{\bar C}}  \left(0.33 \mathrm{I}_{\tens{\bar C}} + \mathrm{II}_{\tens{\bar C}}\right) \left(0.54 \mathrm{II}_{\tens{\bar C}} - 1.34\right) + \mathrm{II}_{\tens{\bar C}} + \log{\left(\frac{\mathrm{II}_{\tens{\bar C}} e^{\mathrm{I}_{\tens{\bar C}}  \left(0.74 \mathrm{I}_{\tens{\bar C}} - \mathrm{II}_{\tens{\bar C}}\right) \log{\left(e^{\sqrt{\mathrm{I}_{\tens{\bar C}}}} + \sqrt{\log{\left(\log{\left(\mathrm{II}_{\tens{\bar C}} \right)} \right)}} \right)}}}{7.98 e^{\mathrm{I}_{\tens{\bar C}}  \left(0.74 \mathrm{I}_{\tens{\bar C}} - \mathrm{II}_{\tens{\bar C}}\right) \log{\left(e^{\sqrt{\mathrm{I}_{\tens{\bar C}}}} + \sqrt{\log{\left(\log{\left(\mathrm{II}_{\tens{\bar C}} \right)} \right)}} \right)}} + 1} \right)}\right) + 0.54 \left(- \sqrt{\mathrm{I}_{\tens{\bar C}}} - \mathrm{II}_{\tens{\bar C}} + \log{\left(\log{\left(\mathrm{II}_{\tens{\bar C}} \right)} \right)}\right) \sqrt{e^{1.04 \sqrt{\mathrm{I}_{\tens{\bar C}} + 0.93 \log{\left(\mathrm{I}_{\tens{\bar C}} \right)}}}}}{\sqrt{e^{1.04 \sqrt{\mathrm{I}_{\tens{\bar C}} + 0.93 \log{\left(\mathrm{I}_{\tens{\bar C}} \right)}}}}}$ & 1.00\\
			3 & 3 & $\bar\Psi_{\rm gMR} = - \mathrm{II}_{\tens{\bar C}}  \left(0.37 \mathrm{II}_{\tens{\bar C}} - 0.63 \exp{\frac{0.14 \mathrm{I}_{\tens{\bar C}} \left(\mathrm{II}_{\tens{\bar C}} - \sqrt{\frac{\exp{\sqrt{\mathrm{II}_{\tens{\bar C}}}}}{\mathrm{I}_{\tens{\bar C}}}}\right)}{\mathrm{II}_{\tens{\bar C}}^{2}}}\right) \left(0.37 \mathrm{I}_{\tens{\bar C}} - 0.94 \mathrm{II}_{\tens{\bar C}} - 0.37 \exp{\frac{0.66 \mathrm{I}_{\tens{\bar C}}}{\log{\left(\mathrm{II}_{\tens{\bar C}} \right)} + 0.13}} - 0.57 \exp{\frac{0.65 \mathrm{I}_{\tens{\bar C}} \left(\mathrm{II}_{\tens{\bar C}} - \sqrt{\frac{\exp{\sqrt{\log{\left(\mathrm{II}_{\tens{\bar C}} \right)}}}}{\mathrm{I}_{\tens{\bar C}}}}\right)}{\mathrm{II}_{\tens{\bar C}}^{2}}} + 3.97\right) $ & 1.00\\		
			3 & 4 & $\bar\Psi_{\rm gMR} = \frac{\mathrm{II}_{\tens{\bar C}}  \left(0.32 \mathrm{I}_{\tens{\bar C}} + 0.32 \mathrm{II}_{\tens{\bar C}} - 1.19\right) \left(0.34 \mathrm{II}_{\tens{\bar C}} - \left(- \mathrm{I}_{\tens{\bar C}} + 1.47 \mathrm{II}_{\tens{\bar C}} + 0.56 \log{\left(\mathrm{I}_{\tens{\bar C}} \right)}\right) \left(\sqrt{\mathrm{II}_{\tens{\bar C}}} - \mathrm{II}_{\tens{\bar C}} + 0.55 \sqrt{\mathrm{I}_{\tens{\bar C}} + 2 \mathrm{II}_{\tens{\bar C}}}\right) + 0.34 \log{\left(\mathrm{I}_{\tens{\bar C}} \right)}\right)}{- \mathrm{I}_{\tens{\bar C}} + 1.47 \mathrm{II}_{\tens{\bar C}} + 0.56 \log{\left(\mathrm{I}_{\tens{\bar C}} \right)}} $ & 1.00\\
			3 & 5 & $\bar\Psi_{\rm gMR} = \frac{\mathrm{II}_{\tens{\bar C}}  \left(1.15 \mathrm{II}_{\tens{\bar C}}  \left(0.23 \sqrt{\mathrm{I}_{\tens{\bar C}}} \mathrm{II}_{\tens{\bar C}} \log{\left(\mathrm{II}_{\tens{\bar C}} \right)} + 0.62 \mathrm{I}_{\tens{\bar C}} \left(\mathrm{I}_{\tens{\bar C}} - 0.52\right)\right) - \left(\mathrm{I}_{\tens{\bar C}} - 0.52\right) \left(0.98 \sqrt{2} \sqrt{\mathrm{I}_{\tens{\bar C}}} \mathrm{II}_{\tens{\bar C}} \sqrt{\log{\left(\mathrm{II}_{\tens{\bar C}} \right)}} + 0.2 \mathrm{I}_{\tens{\bar C}}\right)\right)}{\mathrm{I}_{\tens{\bar C}} - 0.52} $ & 1.00\\
			\bottomrule
		\end{tabular}
		};
	\end{tikzpicture}
	\end{table}

\begin{table}
	\caption{Predicted strain energies for benchmark test using nearly incompressible Mooney-Rivlin model and two different volumetric contributions, where $u(J) = \exp \left({J \left(\log{\left(J \right)} - 1.55\right)}\right)$ for the second function in case 2.}
	\label{tab:gMRPredictionsComp}
	\renewcommand{\arraystretch}{2}
\begin{tabular}{cccc}
	\toprule
	Case & N & Strain energy & $R^2$-score \\
	\midrule
	1 & 1 & $0.63 \ICdev + 0.39 \IICdev - \frac{0.03}{\IICdev} - 9.66 J + 9.49 \exp{J} - 3.64 - 23.0 \exp\left(- \frac{0.73}{J^{2}}\right) $ & 1.00 \\
	1 & 2 & $0.63 \ICdev + 0.39 \IICdev + 25.5 J^{2} - 51.0 J + 25.15$ & 1.00 \\
	1 & 3 & $\frac{0.39 J + \left(J + 0.33\right) \left(0.63 \ICdev + 0.39 \IICdev + 3.06 \log{\left(J^{2} + J + 0.5 \right)}^{3.82}\right) + 19.51}{J + 0.33}$ & 1.00 \\
	1 & 4 & $0.63\ICdev + 0.39 \IICdev + 25.47 J^{2} - 50.98 J + 76.51
	$ & 1.00 \\
	1 & 5 & $0.63 \ICdev + 0.39 \IICdev + 25.5 \left(1.0 J - 1\right)^{2} - 0.34$ & 1.00 \\
	\midrule
	2 & 1 & $0.63 \ICdev + 0.41 \IICdev + \frac{0.03 J}{\ICdev} + \frac{0.12}{\ICdev} + 0.01 \IICdev J + 5.21 J^{2} + 16.6 J + 12.87 + \frac{27.07}{J} $ & 1.00 \\
	2 & 2 & $\begin{aligned}
		&\Big[ 0.02  \left(50.0 u(J) - 3.1\right) \log{\left(\IICdev \exp \left[{- J \left(J \exp \left({J \left(J \log{\left(J \right)} - 0.66\right) \left(\log{\left(J \right)} - 3.28\right)} \right) - 0.67\right)} \right] \right)}\\
		&+\Big( \left(0.65 \ICdev + 1.00J  \left(0.13 \IICdev - 0.57\right)\right) u(J) \Big] \exp \left({- J \left(\log{\left(J \right)} - 1.55\right)}  \right)
	\end{aligned}$ & 1.00 \\
	2 & 3 & $\frac{0.02 \cdot \left(31.47 \ICdev J + J \left(19.53 \IICdev + 19.53 J + 478.53 \left(J - 0.26\right)^{2} - 1.29\right) + 1150.37 \left(0.61 J + 1\right)^{2}\right)}{J}$ & 1.00 \\
	2 & 4 & $\left(0.11 \IICdev + 32.57\right) \left(\log{\left(J \right)}^{2} + \log{\left(\ICdev + 39.9 \right)}\right) - 0.9 \log{\left(\ICdev + 2.99 \right)}$ & 1.00 \\
	2 & 5 & $0.63 \ICdev + 0.39 \IICdev + 6.39 \exp{J} + 6.6 + \frac{20.61}{\left(0.5 \left(0.75 \log{\left(\log{\left(J \right)} + 0.65 \right)} + 1\right)^{2} + 1\right)^{2}}$ & 1.00 \\
	\bottomrule
\end{tabular}
\end{table}

\begin{table}
	\caption{Recovered damage models for benchmark test using Ogden-Roxburgh model.}
	\label{tab:OR}
	\renewcommand{\arraystretch}{2}
	\begin{tabular}{ccc}
		\toprule
		N & Damage function & $R^2$-score \\
		\midrule
		1 & $ \eta \left(\Psi, \Psi_{\text{max}}\right) = 0.83\erf\left(\frac{4.0 \Psi - 4.0\Psi_{\text{max}}}{2\Psi_{\text{max}} + 8.0}\right) + 1.0$ & 1.00 \\
		2 & $ \eta \left(\Psi, \Psi_{\text{max}}\right) = 0.83\erf\left(\frac{4.0 \Psi - 4.0 \Psi_{\text{max}}}{2 \Psi_{\text{max}} + 8.0}\right) + 1.0$ & 1.00 \\
		3 & $ \eta \left(\Psi, \Psi_{\text{max}}\right) = 0.83\erf\left(\frac{\Psi - \Psi_{\text{max}}}{0.5\Psi_{\text{max}} + 2.0}\right)\erf\left(0.07\Psi + 1.21\Psi_{\text{max}} + 3.09\right) + 1.0$ & 1.00 \\
		4 & $ \eta \left(\Psi, \Psi_{\text{max}}\right) = 0.83\erf\left(\frac{2.0\Psi - 2.0 \Psi_{\text{max}}}{\Psi_{\text{max}} + 4.0}\right) + 1.0$ & 1.00 \\
		5 & $ \eta \left(\Psi, \Psi_{\text{max}}\right) = 1.0 - 0.83\erf\left(1.98\frac{-\Psi + \Psi_{\text{max}}}{\Psi_{\text{max}} + 4.0}\right)$ & 1.00 \\
		\bottomrule
	\end{tabular}
\end{table}
\clearpage

\printcredits


\bibliography{cas-refs} 
\bibliographystyle{unsrt}

\end{document}